\documentclass[aps,prl,twocolumn,showpacs,lengthcheck,reprint]{revtex4-1}
\usepackage{times}
\usepackage{natbib}
\usepackage{hyperref}
\usepackage{amsfonts}
\usepackage{amsmath}
\usepackage{amssymb}
\usepackage{graphicx}%
\providecommand{\U}[1]{\protect\rule{.1in}{.1in}}

\usepackage{color}

\begin{document}

\title{Chladni solitons and the onset of the snaking instability for dark 
solitons in confined superfluids}
\author{A. Mu\~{n}oz Mateo}
\email{A.M.Mateo@massey.ac.nz}
\author{J. Brand}
\email{J.Brand@massey.ac.nz}
\affiliation{Centre for Theoretical Chemistry and Physics and New Zealand Institute for
Advanced Study, Massey University, Private Bag 102904 NSMC, Auckland 0745, New Zealand}

\date{\today}

\pacs{3.75.Lm,67.85.De,67.85.Lm}

\begin{abstract}
Complex solitary waves composed of intersecting vortex lines are 
predicted in a channeled superfluid.
Their shapes in a 
cylindrical trap include a cross, spoke wheels, and Greek $\Phi$, and trace the 
nodal lines of unstable vibration modes of a planar dark soliton in analogy to 
Chladni's figures of membrane vibrations.
The stationary solitary waves extend a family of solutions that include the 
previously known solitonic vortex and vortex rings. Their bifurcation points 
from the dark soliton indicating the onset of new unstable modes of the snaking 
instability are predicted from scale separation for Bose-Einstein 
condensates (BECs) and superfluid Fermi gases across the BEC--BCS crossover, 
and confirmed by full numerical calculations.   Chladni solitons could be observed in ultra-cold gas experiments by seeded decay of dark solitons.
\end{abstract}

\maketitle



Solitons are the hallmark of nonlinear physics. They are ubiquitous in one-dimensional wave propagation and appear in diverse systems ranging from carbon nanotubes 
\cite{Chamon2000,Deshpande2008}  and water waves \cite{Chabchoub2013} to nonlinear optics \cite{Kivshar1998,Amo2011} and superfluid atomic gases \cite{Burger1999,Denschlag2000,Yefsah}. Experimental signatures of vortex rings in trapped BECs \cite{Anderson2001,Dutton2001,ginsberg05} and the recent observations of solitonic vortices  in atomic BECs  \cite{Becker2013,Donadello2014} and a superfluid Fermi gas \cite{Yefsah,ku2014} highlight the richness of solitary wave motion in waveguide-like geometries beyond the one-dimensional paradigm. A solitonic vortex consists of a single vortex line crossing the transverse diameter of an elongated, trapped condensate \cite{brand01a,Brand2002,Komineas2003}, while the vortex line forms a closed loop in a vortex ring. In contrast to vortex lines \cite{pitaevskii03:book} and rings \cite{Jones1982} in bulk superfluids, both structures are heavily influenced by the waveguide trap, leading to localisation with a well-defined step of the superfluid phase across the excitation \cite{brand01a,Komineas2002,Komineas2003a} and modified dynamical properties \cite{fetter01:vortices,ku2014,Horng2006,Hsueh2007,Pitaevskii2013,Bulgac2013a}.  The systematic numerical study of solitary waves in wave-guide-trapped superfluids has initially been restricted to axisymmetric solitary waves in BECs at weak nonlinearities \cite{Komineas2002,Komineas2003} and has later included the non-axisymmetric solitonic vortex \cite{Komineas2003a}. Here we seek to close the apparent gap and consider complex solitary waves with crossing vortex lines and broken axial symmetry.

Solitary waves probe superfluids at the mesoscopic length scale of the healing length. The physics at these length scales is not very well understood for strongly-correlated superfluids, such as the  resonant atomic Fermi gas across the BEC--BCS crossover \cite{Giorgini2008}. In order to provide a reference point for further study, we here employ the recently developed coarse-grained version of the Bogoliubov-de Gennes equation \cite{Gennes1999} for the crossover superfluid by Simonucci and Strinati \cite{Simonucci2014} at zero temperature, which includes the Gross-Pitaevskii 
equation in the BEC limit. Previously, the Bogoliubov-de Gennes equations have been used to describe stationary \cite{Antezza2007} and moving dark solitons \cite{Scott2011,Spuntarelli2011,Liao11pr:FermiSolitons,Scott2012}, and the snaking instability \cite{Cetoli2013}  in a bulk fermionic superfluid. A variety of other approaches has also been used to simulate dynamics in the crossover Fermi superfluid \cite{Bulgac2011a,Bulgac2014,Wen2013,Scherpelz2014}. 



\begin{figure}[tb]
\includegraphics[width=8.5cm]{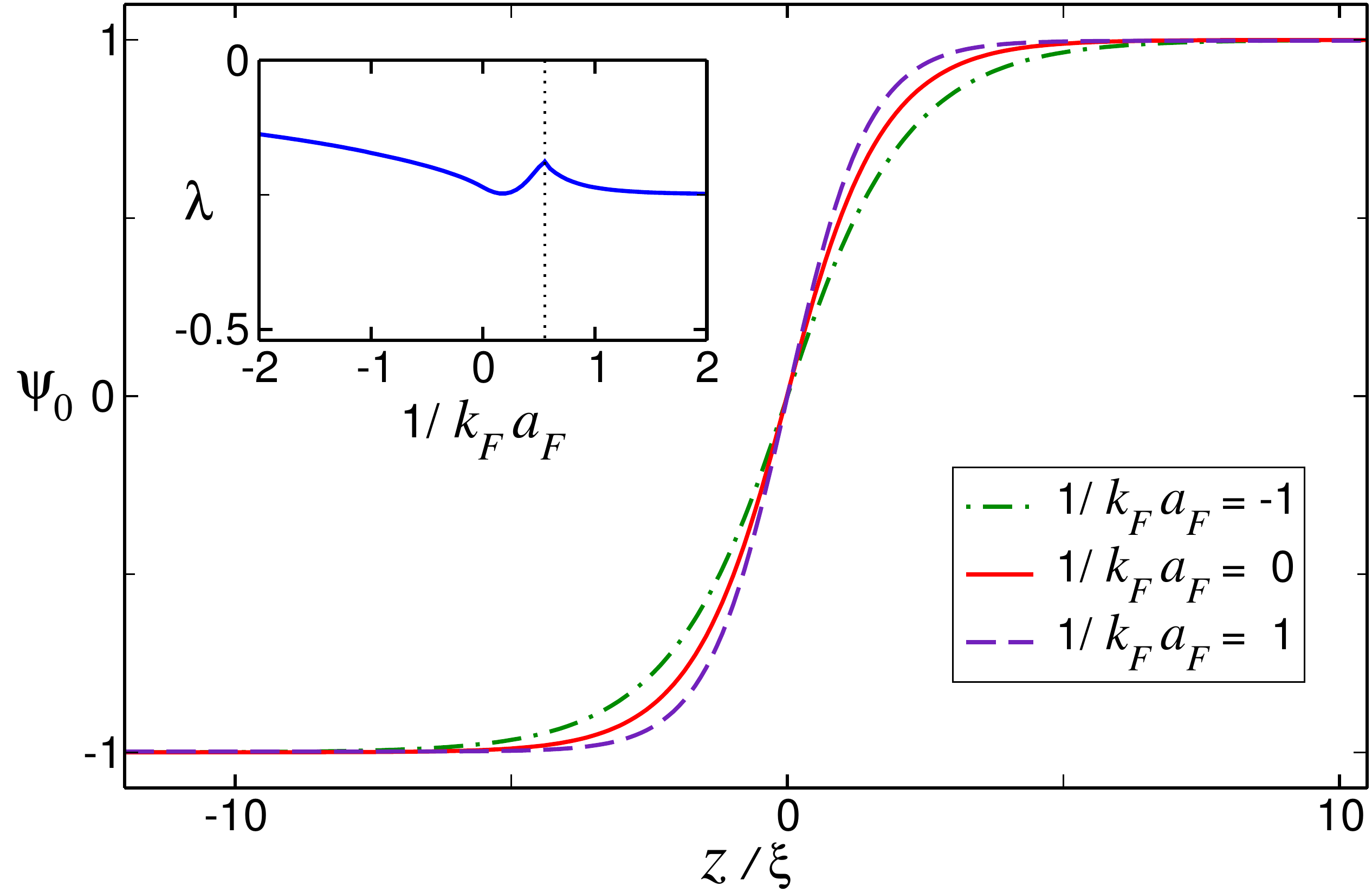}
\caption{ Superfluid order parameter $\psi_0(z/\xi)$ of
kink solutions of Eq.~(\ref{eq:nls}) in the BEC--BCS crossover for different values of the coupling 
constant $\eta=1/k_Fa_F$, where the axial coordinate $z$ is measured in units of the healing 
length $\xi=\hbar/\sqrt{4m\mu^B}$. The inset shows the (bound) ground state 
eigenvalue $\lambda$ of Eq.~(\ref{eq:u1}), which modulates the energy available 
for the excitation of transverse modes. The dotted line indicates the change of 
sign of the chemical potential.}
\label{fig:kink}
\end{figure}

The conceptually simplest solitary wave in a superfluid with a scalar, complex order parameter is a stationary kink, or dark soliton, characterised by a node in the order parameter and exponential healing to the background fluid \cite{Tsuzuki1971} (see also Fig.\ \ref{fig:kink}). In two or three dimensions, however, the kink is unstable towards snaking \cite{Kuznetsov1988} and its stability can only be restored by constricting 
the geometry to a narrow channel, as e.g.\ in an elongated atom trap 
\cite{Muryshev1999}. The appearance of instability modes is connected with the 
bifurcation of new solitary waves \cite{Brand2002}. In this work, we study the 
complete family of stationary solitary waves bifurcating from the dark soliton 
in a wave-guide trap geometry. We show that each 
new solitary wave at the point of bifurcation corresponds to a membrane 
vibration mode 
with vortex filaments along its nodal lines. They correspond to the famous 
Chladni figures that visualise the nodal lines of plate vibrations 
\cite{Chladni1787}. We thus call the whole family \emph{Chladni solitons}. 
Within a scale-separating approximation we derive a simple formula 
[Eq.~(\ref{eq:bif})] for the bifurcation points in a cylindrically trapped gas 
in terms of two integer quantum numbers $p$ and $l$, describing the number of 
vortex rings and radially extending vortex lines, respectively. These results 
are corroborated by full numerical solutions of the Gross-Pitaevskii equation. 
We suggest strategies for observing higher order Chladni solitons through 
the decay of dark solitons.

\emph{Nonlinear Schr\"odinger model for the BEC--BCS crossover} -- 
In the approach of Simonucci and Strinati \cite{Simonucci2014}, 
the superfluid order parameter $\Delta(\mathbf{r})$ at zero 
temperature is determined as 
the self-consistent solution of a nonlinear Schr\"odinger equation, which we 
write as
\begin{align} \label{eq:lpda}
&\left[-\frac{\hbar^{2}}{4m}\nabla^{2} + f(\mu_\mathrm{loc}(\mathbf{r}),|\Delta(\mathbf{r})|)\right] \Delta(\mathbf{r})=0 , 
\end{align}
where $m$ is the fermion mass and the local chemical potential $\mu_\mathrm{loc}(\mathbf{r}) = \mu - V(\mathbf{r})$ combines the chemical potential $\mu$ and the trapping potential $V(\mathbf{r})$.
The function $f$ provides the nonlinearity and can be written as
\begin{align} \label{eq:f}
f(\mu,|\Delta|) = -4\mu +4 |\Delta| \frac{I_{5}}{I_{6}} - \frac{\pi \hbar \sqrt{|\Delta|}}{\sqrt{2m} a_{F} I_{6}} ,
\end{align}
where $a_{F}$ is the $s$-wave scattering length of fermions. The functions
$I_{5}(x_{0})$ and $I_{6}(x_{0})$ are expressed by elliptic 
integrals and evaluated with the argument $x_{0} = \mu/|\Delta(\mathbf{r})|$ 
\footnote{\label{fn:Elliptic}In terms of the complete elliptic integrals of the 
first and second kind $K(\kappa)$ and $E(\kappa)$, respectively, 
$I_{5}(x_{0})=(1+x_{0}^{2})^{1/4} E(\kappa) 
-K(\kappa)/[4x_{1}^{2}(1+x_{0}^{2})^{1/4}]$ and 
$I_{6}(x_{0})=K(\kappa)/[2(1+x_{0}^{2})^{1/4}]$, with 
$x_{1}^{2}=(\sqrt{1+x_{0}^{2}}+x_{0})/2$ and $\kappa^2 = 
x_{1}^{2}/\sqrt{1+x_{0}^{2}}$ \cite{Simonucci2014,Marini1998}. We use notation 
for elliptic integrals conforming with \cite{Olver2010}.}.  Equation 
(\ref{eq:lpda}) approximates the Bogoliubov-de Gennes equation 
\cite{Gennes1999,Leggett1980a}. For homogeneous order parameter  $\Delta_{0}>0$, 
 the resulting equation $f(\mu,\Delta_{0})=0$ together with the accompanying 
equation for the density \cite{Simonucci2014} is equivalent to standard BCS theory for 
the crossover superfluid \cite{Eagles1969,Leggett1980a,Marini1998}.
In the BEC limit $a_{F} \to 0+$, Eq.~(\ref{eq:lpda}) reduces to the
Gross-Pitaevskii equation 
\footnote{For large and positive $1/(k_{F}a_{F}) \gg 1$ (BEC limit), Eq.\
(\ref{eq:lpda}) reduces to the Gross-Pitaevskii equation
\begin{align} \label{eq:GPE}
\left[-\frac{\hbar^{2}}{2m_{B}}\nabla^{2} -\mu^B + V_{B}(\mathbf{r}) + g_{B}|\phi|^{2}\right] \phi(\mathbf{r})=0 ,
\end{align}
where $m_{B}=2m$ is the mass of bosons,
$V_{B}=2V$, $\mu^B= 2\mu + \hbar^2 /(m a_{F}^{2})$ is the boson chemical 
potential, 
$\phi(\mathbf{r})=\Delta(\mathbf{r}) \sqrt{\frac{m^{2} a_{F}}{8 \pi \hbar^{4}}}$ 
the Gross-Pitaevskii order parameter, and $g_{B}= 4 \pi \hbar^{2} a_{B}/m_{B}$ the coupling 
strength  \cite{Simonucci2014}.}.

We now proceed to solve Eq.\ (\ref{eq:lpda}) for solitary waves by separation of
scales, where the characteristic length scale for the solitary waves $\xi$ is 
assumed much smaller than the length scale of order parameter variation due to 
the presence of the trapping potential $V(\mathbf{r})$, on which spatial 
derivatives can be neglected. This will allow us to understand the solitary 
waves as excitations on top of a background order parameter 
$\Delta_{0}(\mathbf{r})$. The latter is obtained using the Thomas Fermi 
approximation $f(\mu_\mathrm{loc},\Delta_{0})=0$, which results from Eq.\ 
(\ref{eq:lpda}) by neglecting the spatial derivative term. We introduce the 
length scale $\xi(\mathbf{r})$ locally, which we expect to be relevant for the 
solitary waves, by
\begin{align} \label{eq:xi}
\frac{\hbar^2}{4 m \xi^2(\mathbf{r})} = \mu^B_\mathrm{loc}(\mathbf{r}) \equiv  
2\mu_\mathrm{loc}(\mathbf{r}) +\sigma\frac{\hbar^2}{m a_{F}^{2}} ,
\end{align}
where $\mu^B_\mathrm{loc}(\mathbf{r})$ is a local bosonic chemical potential 
and $\sigma =1$ for $a_F > 0$ and $\sigma=0$ otherwise. Rescaling 
Eq.\ (\ref{eq:lpda}) with this length scale, normalising the order parameter 
$\psi(\tilde{\mathbf{r}}) = \Delta(\tilde{\mathbf{r}} 
\xi)/\Delta_{0}(\tilde{\mathbf{r}} \xi)$ and ignoring spatial derivatives of the 
slowly-varying $\xi(\mathbf{r})$ consistent with the Thomas Fermi 
approximation, we obtain 
\begin{align} \label{eq:nls}
-\frac{1}{2} \tilde{\nabla}^2 \psi -\psi + g(|\psi|^2) \psi = 0 ,
\end{align}
where 
$g(|\psi|^2) = 1+ f(\mu_\mathrm{loc},\Delta_0|\psi|)/{2\mu^B_\mathrm{loc}}$.
In units of the background fermion density $n_0$ with the Fermi wavenumber
$k_F = (3 \pi^2 n_0)^{1/3}$ and the Fermi energy $E_F = \hbar^2k_F^2/(2 m)$ 
\footnote{The background density $n_0$ used here to define the unit system is 
related to the chemical potential $\mu^B_\mathrm{loc}$ and $a_F$ by the standard 
relations of crossover BCS theory \cite{Simonucci2014}. It depends on position 
weakly, i.e.\ its spatial derivatives are ignored.}, the function $g$ can be 
expressed with $\tilde{\mu}= \mu/E_F$ and the coupling strength $\eta = 1/(k_F 
a_F)$ as
\begin{align}
g(|\psi|^2) = \frac{\sigma \eta^2 + \tilde{\mu} \frac{I_5}{x_0 I_6} - \frac{\eta\pi \sqrt{\tilde{\mu}}}{4 \sqrt{x_0} I_6}}{\tilde{\mu} + \sigma\eta^2} ,
\end{align}
where $I_5$ and $I_6$ are evaluated at $x_0 = x_0^\mathrm{bg}/|\psi|$ and
$x_0^\mathrm{bg}=\mu_\mathrm{loc}/\Delta_0$ is  the background value 
\cite{Note1}. The dimensionless function $g$ defines a generalised nonlinearity 
in the nonlinear Schr\"odinger equation (\ref{eq:nls}). It weakly depends on 
position through dependence of $\eta$ on the background density 
except in the BEC limit and at unitarity, where it is position independent.
While the BEC limit $\eta\to +\infty$ [where $g(|\psi|^2) \to (1+ |\psi|^2)/2$] 
recovers the cubic nonlinear Schr\"odinger equation, the function $g$ simplifies 
at unitarity ($\eta =0$) to $g(|\psi|^2) = I_5/(x_0 I_6)$, where $x_0 = 
x_0^u/|\psi|$ and $x_0^u \approx 0.8604$ is the background value at unitarity.


\emph{Kink solution and its bifurcations} -- The nonlinear Schr\"odinger
equation (\ref{eq:nls}) supports a real kink solution 
$\psi_0(\tilde{\mathbf{r}})$ that depends on $\tilde{z}$ only with a single node 
at $\tilde{z}=0$ and boundary condition $\psi_0 \to \pm 1$ as $\tilde{z} \to 
\pm \infty$, respectively, as shown in Fig.\ \ref{fig:kink}. 
Bifurcations of symmetry breaking solutions are found by 
using the ansatz $\psi = \psi_0+ i \delta \psi$ and linearising Eq.\ 
(\ref{eq:nls}) for a completely imaginary perturbation $i\delta \psi$. Due to 
the symmetry of the original kink solution $\psi_0$, the perturbation can 
further be separated with the product ansatz $\delta\psi 
=\chi(\tilde{x}\xi,\tilde{y}\xi)u(\tilde{z})$. For $u$ this yields the 
eigenvalue equation
\begin{align}\label{eq:u1}
\left[-\frac{1}{2}\partial_{\tilde{z}\tilde{z}} -1 + g(|\psi_0(\tilde{z})|^{2})\right] u &= \lambda u . 
\end{align}
This Schr\"odinger equation with localised potential well has a one-node 
solution $u=\psi_0$, which corresponds to the Goldstone mode of phase symmetry 
with
$\lambda=0$. More relevant for bifurcating solitary waves, Eq.\ (\ref{eq:u1}) 
has a node-less and localised solution with eigenvalue $\lambda<0$. In the BEC 
limit, $1-g(|\psi_0|^2) = \mathrm{sech}^2(\tilde{z}/\sqrt{2})/2$ is a 
Rosen-Morse 
potential with analytically known solution $u = 
\mathrm{sech}(\tilde{z}/\sqrt{2})$ and 
eigenvalue $\lambda = -1/4$ \cite{Rosen1932}. For finite values of the coupling 
constant $\eta$, $\lambda$ can be determined numerically and is shown in the inset of 
Fig.~\ref{fig:kink}. At unitarity, we obtain $\lambda=-0.24$.

\begin{figure}[tb]
\includegraphics[width=8.5cm]{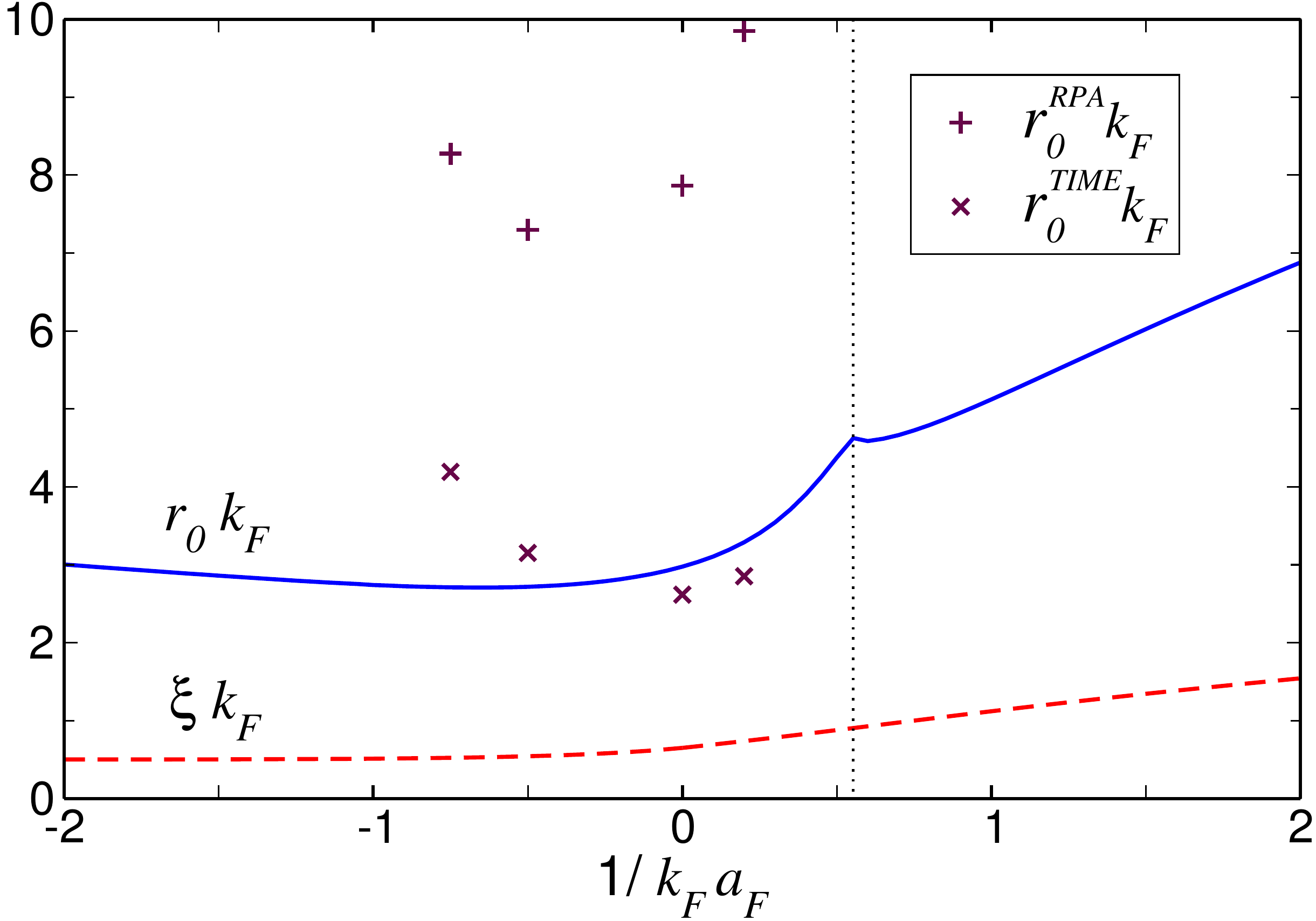}
\caption{Length scale $r_{0}$ (full line) of the snaking instability given by Eq.\ 
(\ref{eq:r0}) and healing length $\xi$ of Eq.~(\ref{eq:xi}) (dashed line) as 
functions of the interaction parameter $1/k_Fa_F$ of the BEC--BCS crossover. 
For comparison, the results  from time-dependent  Bogoliubov-de Gennes 
simulations and RPA of Ref.~\cite{Cetoli2013} are shown. }
\label{fig:r0}
\end{figure}

For the transverse wave function $\chi$  we obtain
by restoring the unscaled coordinates the wave equation 
\begin{align}
\label{eq:chi1}
\left[ -\frac{\xi^2}{2}\nabla_{\perp}^{2}+ \lambda\right]\chi &=0 .
\end{align}
In the absence of a trapping potential, $\xi$ is constant. The real-valued solutions are readily found as standing 
waves with nodal lines separated by a distance 
\begin{align}
\label{eq:r0}
r_0 =  \frac{\pi}{\sqrt{-2\lambda}}  \xi,
\end{align}
which is the minimum length scale of the snaking instability 
\cite{Brand2002}. In Fig.~\ref{fig:r0} the value of $r_0$ is 
compared with the numerical results of Ref.~\cite{Cetoli2013}, which suffered 
inaccuracies due to numerical limitations for grid size and cutoff parameters.

In the presence of a trapping potential, the influence of the snaking 
instability can be made explicit in rewriting Eq.\ (\ref{eq:chi1}) as
\begin{align}
\label{eq:chi2}
\left\{ -\frac{\hbar^2}{4m}\nabla_{\perp}^{2} - 
\lambda\left[4V(\mathbf{r}) -2\mu^B\right]\right\}\chi &=0 .
\end{align}
This is a two-dimensional Schr\"odinger equation with a scaled external 
potential and prescribed eigenvalue. 
For the case of 
harmonic isotropic trapping with $V(\mathbf{r})=\frac{1}{2}m 
\omega_{\perp}^2(x^2+y^2)$ we obtain the condition for the bifurcation points
\begin{align} \label{eq:bif}
 \frac{\mu^B}{\hbar \omega_{\perp}} = \frac{2p +l +1}{\sqrt{-2\lambda}},
\end{align}
where $p$ and $l$ are the radial and azimuthal quantum numbers of the cylindrical harmonic 
oscillator eigenfunctions $|\,p,l\rangle\,
\propto\exp(-r_\perp^2/2a_\perp^2)r_\perp^lL^l_p(r_\perp^2/a_\perp^2)
\cos(l\theta)$, and $L^l_p$ are the associated Laguerre polynomials in 
$r_\perp^2=x^2+y^2$, 
with the characteristic oscillator length $a_{\bot}=\sqrt{\hbar/m\omega_\bot}$. 
The condition (\ref{eq:bif}) together with the harmonic oscillator
eigenfunctions defines the bifurcation points as well as the shape, symmetry, 
and degeneracy of the corresponding bifurcating vortex solutions. 

\begin{figure}[tb]
\includegraphics[width=8.5cm]{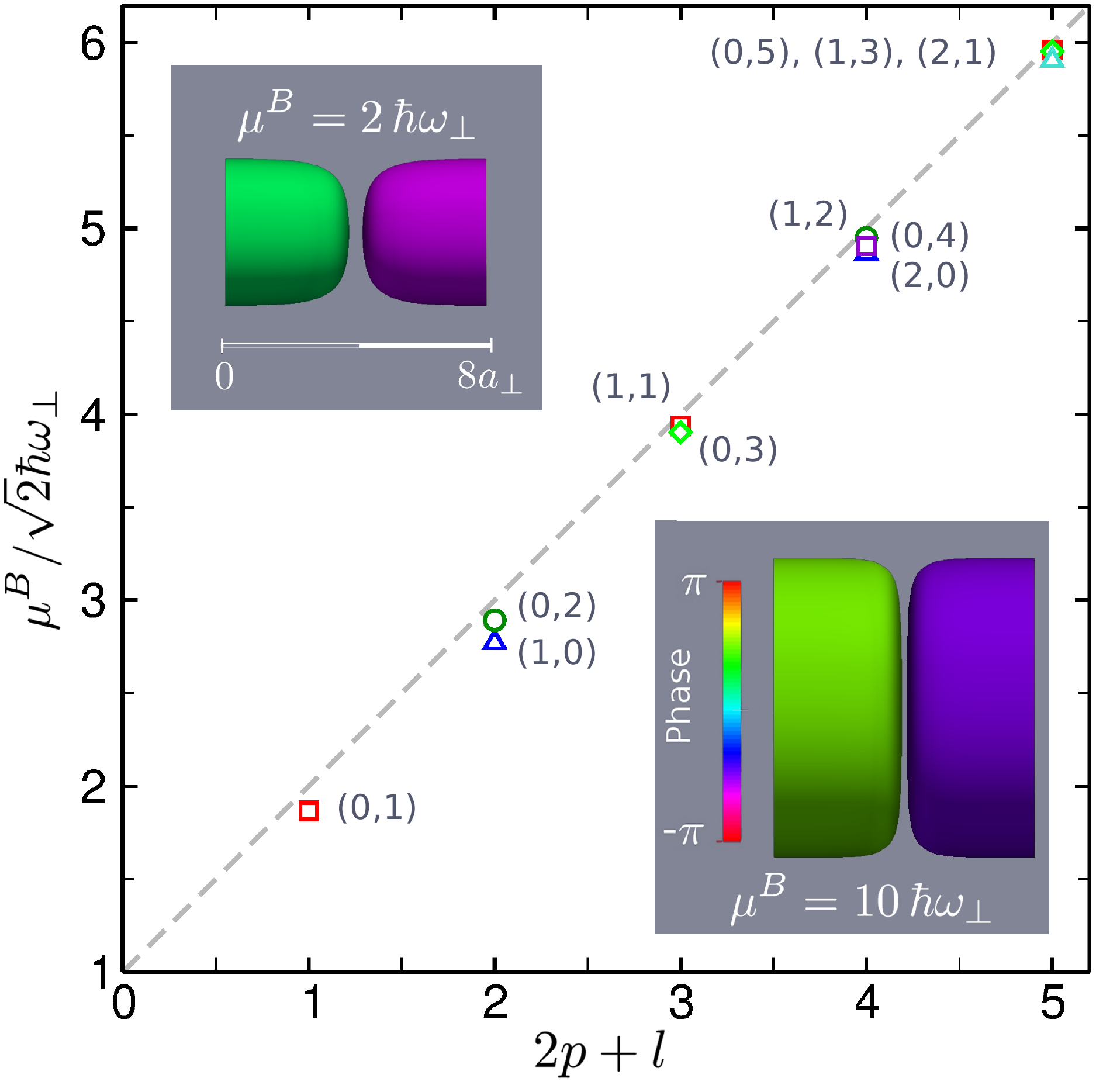}
\caption{ Bifurcation points for the emergence of 
Chladni solitons with quantum numbers $(p,l)$ from the planar kink in a cylindrically confined BEC.
Numerical data (labeled open symbols) obtained from solving the 
Gross-Pitaevskii equation are compared with the  analytical prediction (dashed 
line) of Eq.\ (\ref{eq:bif}) with $\lambda= -\frac{1}{4}$. The insets display 3D 
density isocontours of dark soliton states at 5$\%$ of maximum density in the 
weak (upper left) and strong (lower right) nonlinearity regimes.}
\label{Fig3}
\end{figure}

\begin{figure}[tb]
\includegraphics[width=8.5cm]{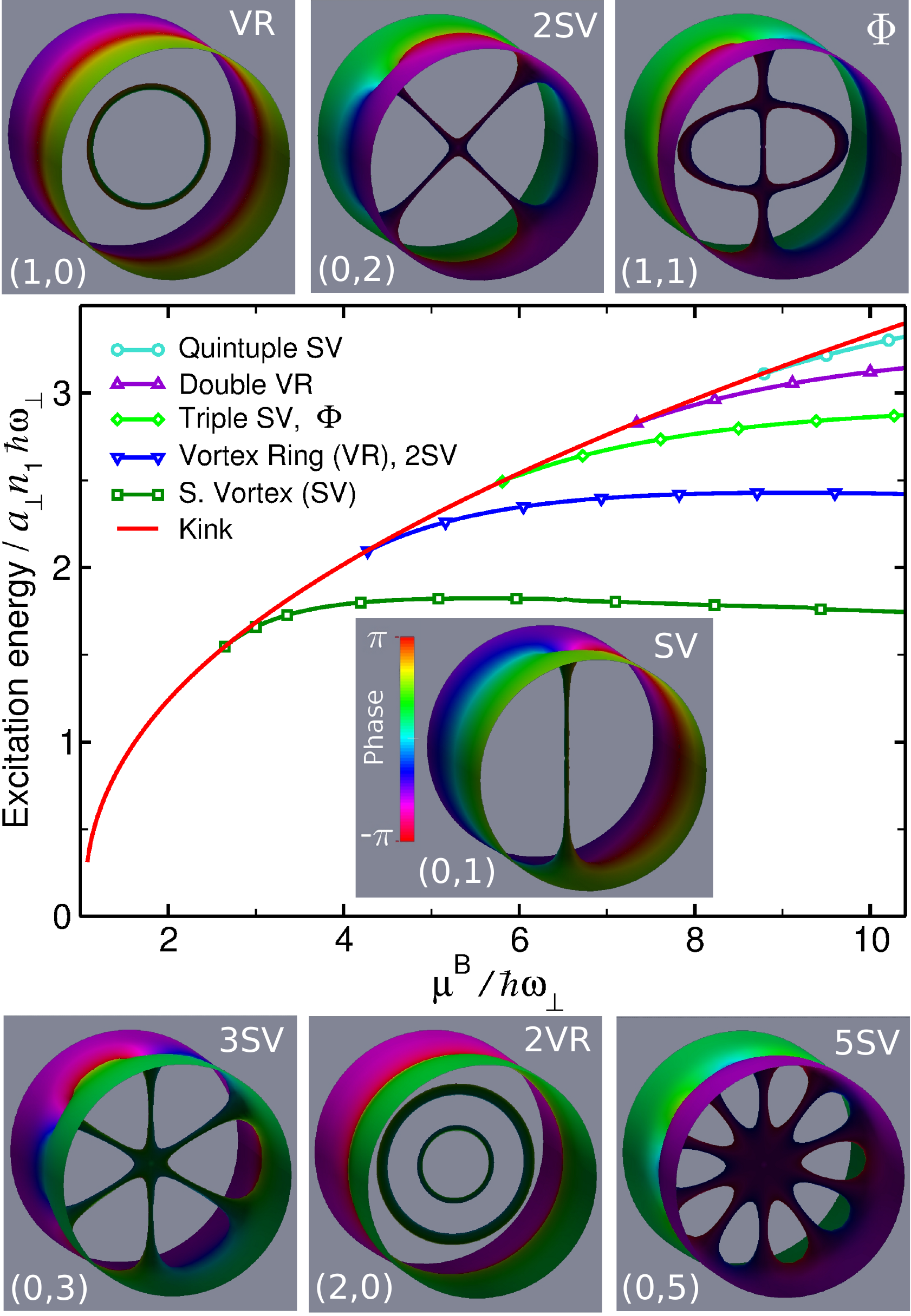}
\caption{Free excitation energy $F_{pl}$ 
of 
stationary solitary waves vs.\ chemical potential in a cylindrically trapped BEC 
(central panel). The full (red) line corresponds to the dark soliton [see insets 
in Fig. (\ref{Fig3})] and lines with symbols correspond to Chladni solitons  
$(p,l)$ originated from the bifurcation points of 
Fig.~\ref{Fig3}.
Units of the transverse harmonic trap and axial density $n_1$ are used as shown.
The insets show density isocontours at 5$\%$ of maximum density for the 
different Chladni solitons with $\mu^B=10\hbar\omega_{\bot}$.}
\label{Fig4}
\end{figure}

Numerical data for the bifurcation points in the BEC limit are shown in 
Fig.~\ref{Fig3}.
The agreement with Eq.\ (\ref{eq:bif}) is surprisingly good even for 
small quantum numbers. The first bifurcation point corresponds to a single 
straight nodal line ($p=0, l=1$) and leads to the solitonic vortex (SV), shown 
in the central inset of Fig.~\ref{Fig4}. It marks the onset of the snaking 
instability for the kink. Its accurate numerical value of $\mu^B /\hbar 
\omega_\perp= 2.65$ is very close to the value $2\sqrt{2}\approx 2.82$ of Eq.\ 
(\ref{eq:bif}) and consistent with Ref.~\cite{Komineas2003}.
The next bifurcation point comes from the 
degenerate solutions
$(1,0)$ and $(0,2)$, 
corresponding to a single vortex ring (VR) originating from a radial node and 
to two crossed solitonic vortices (2SV) originating from two azimuthal nodes,
respectively. Our numerical results resolve the  degeneracy breaking between the VR ($\mu^B /\hbar \omega_\perp= 3.9$) and 2SV  ($4.1$)
bifurcation points (see Fig.~\ref{Fig3}) that cannot be captured by our 
analytical model ($3\sqrt{2}\approx 4.2$). Similar degeneracy breaking occurs 
for higher quantum numbers $(p,l)$. Overall, the agreement with the analytic 
result of  Eq.\ (\ref{eq:bif}) improves for higher quantum numbers corresponding 
to larger $\mu$. This is expected since the  underlying 
approximation -- scale separation between Thomas Fermi radius and healing length 
-- is increasingly satisfied.

In the more general case of an anisotropic transverse trapping potential $V(\mathbf{r})=\frac{1}{2}m(\omega_x^2x^2 +\omega_y^2 y^2)$ the bifurcation condition (\ref{eq:bif}) is modified to read $\mu^B\sqrt{-2\lambda}=(n_x+\frac{1}{2})\hbar\omega_x + (n_y+\frac{1}{2})\hbar\omega_y$. The corresponding harmonic oscillator eigenfunctions are given by the well known Hermite functions and thus the symmetry of the corresponding Chladni solitons changes. For small anisotropies, we expect a continuous transition from Laguerre to Hermite shapes as is shown by Ince polynomials \cite{Bandres2004}.

\emph{Chladni solitons} --
We have numerically determined the solitary wave solutions originating from the
bifurcation points of Fig.~\ref{Fig3}  up to $\mu^B = 10 \hbar\omega_\perp$ 
using a Newton-Raphson scheme for the Gross-Pitaevskii equation. The results are 
summarised in Fig.~\ref{Fig4}, where the free excitation energies 
$F_{pl}=E_{pl}-\mu^B N_{pl}-(E_0-\mu^B N_0)$ are measured relative to the 
ground state $\phi_0$ of Eq.\
(\ref{eq:GPE}) \footnote{We use the standard  expressions for energy
\begin{align}
E = \int d\mathbf{r} \left\{\frac{\hbar^2}{2 m_B} |\nabla\phi|^2 + V_B|\phi|^2 + \frac{g_B}{2}|\phi|^4 \right\},
\end{align}
and particle number $N=\int d\mathbf{r}\, n$,
consistent with the Gross-Pitaevskii equation \cite{Note2}, where $n(\mathbf{r})=|\phi(\mathbf{r})|^2$.}. 
For every tuple of quantum numbers $(p,l)$ we obtain solitary wave solutions with symmetry and degeneracy consistent with the solutions of Eq.\ (\ref{eq:chi2}), which is maintained  for $\mu^B$ above the bifurcation point. I.e.\ Chladni solitons involving vortex rings have a discrete 2-fold degeneracy while all solitons involving radial vortex lines have a continuous degeneracy corresponding to azimuthal rotation.

In addition to the previously known dark soliton (kink), solitonic vortex
$(0,1)$, and vortex ring $(2,0)$, more complex Chladni solitons comprise 
\emph{spoke wheels} $(0,l)$ consisting of $l>1$ intersecting radial vortex 
lines, multiple nested vortex rings $(p,0)$ and the $\Phi$-type soliton $(1,1)$, 
which is the simplest soliton with intersecting vortex ring(s) and radial 
line(s). Note that Chladni solitons originating from near-degenerate bifurcation 
points, e.g.\ the 2SV cross $(0,2)$ and vortex ring $(1,0)$ are almost 
degenerate also for larger $\mu^B$ and their small energy difference is not 
resolved at the scale of Fig.\ 
\ref{Fig4}. The same is true for the higher near-degenerate branches.

Our numerical simulations indicate that Chladni solitons with intersecting
vortex lines could be observed in cigar-shaped BECs. Although the solitonic 
vortex $(0,1)$ is the only dynamically stable Chladni soliton for $\mu^B$ beyond 
the first bifurcation point of Eq.~(\ref{eq:bif}), the expected lifetimes of the 
$2p+l>1$ Chladni solitons are comparable to those of the already observed vortex 
rings. Detailed  stability studies will be published elsewhere \footnote{A. 
Mu\~noz Mateo and J. Brand, in preparation (2014).}. 
Complex Chladni solitons can be prepared in an atomic BEC by appropriately
seeding the snaking instability of a dark soliton.
For this purpose, the previously determined wave function $\psi = 
\psi_0+i\delta\psi$ at the bifurcation point, should be an excellent initial 
configuration, since the infinitesimal part $\delta\psi$ develops into a 
dynamically unstable mode leading to the desired complex soliton for $\mu^B$ 
larger than the critical value at bifurcation \cite{Brand2002}.
%
As a first step we propose to prepare a zero velocity dark-bright soliton in a
two-component BEC \cite{Busch2001} as in Refs.~\cite{Anderson2001,Becker2008}, 
where the nodal plane of a kink in component $|1\rangle$  is filled with a 
phase-coherent atomic BEC of a second hyperfine component $|2\rangle$. In a 
second step, the bright component $|2\rangle$ (or a small part of it) is 
transferred to state $|1\rangle$ with the appropriate phase pattern of the 
solution $\delta\psi=\chi u$ of Eqs.~(\ref{eq:u1}) and (\ref{eq:chi2}), shifted 
by $\frac{\pi}{2}$ compared to the kink solution. This  could be realised 
following \cite{Dum1998} by transferring the phase pattern of a focused axial 
Laguerre-Gaussian laser beam using a two-photon Raman transition. Finally, any 
remaining $|2\rangle$ atoms are removed \cite{Anderson2001}.

\emph{Acknowledgements} -- We are grateful to Michael Bromley and Xiaoquan Yu for useful discussions.

\bibliography{Solitons,Fermi_Gas,trapped_dark_soliton_7}

\end{document}